\documentclass[aps,draft,preprintnumbers,amsmath,amssymb,footinbib]{revtex4}
\usepackage[latin9]{inputenc}
\usepackage{textcomp}
\usepackage{mathtools}
\usepackage{amsmath}
\usepackage{amssymb}
\usepackage{esint}

\makeatletter

\DeclareFontEncoding{LGR}{}{}

\ProvideTextCommand{\~}{LGR}[1]{\char126#1}

\@ifundefined{textcolor}{}
{%
 \definecolor{BLACK}{gray}{0}
 \definecolor{WHITE}{gray}{1}
 \definecolor{RED}{rgb}{1,0,0}
 \definecolor{GREEN}{rgb}{0,1,0}
 \definecolor{BLUE}{rgb}{0,0,1}
 \definecolor{CYAN}{cmyk}{1,0,0,0}
 \definecolor{MAGENTA}{cmyk}{0,1,0,0}
 \definecolor{YELLOW}{cmyk}{0,0,1,0}
}

\usepackage[final]{graphics}
\usepackage{amsfonts}
\usepackage{color}
\usepackage{MnSymbol}

\def\be{\begin{equation}}
\def\ee{\end{equation}}
\def\bea{\begin{eqnarray}}
\def\eea{\end{eqnarray}}

\def\beq{\begin{equation}}
\def\eeq{\end{equation}}

\renewcommand{\v}[1]{ \ensuremath{  \underline {#1} }}

\makeatother

\begin{document}

\title{Systematics of angular correlations in 2 gluon production: dependence
on the projectile geometry } 

\title{Does  shape matter? $v_2$ vs eccentricity in small x gluon production} 

\preprint{CERN-TH-2018-077, RBRC-1273}
\author{Alex Kovner}
\affiliation{Physics Department, University of Connecticut, 2152 Hillside Road, Storrs, CT 06269, USA}
\affiliation{Theoretical Physics Department, CERN, CH-1211 Geneve 23, Switzerland}

\author{Vladimir V. Skokov}
\affiliation{RIKEN/BNL, Brookhaven National Laboratory, Upton, NY 11973, USA}

\date{\today}
\begin{abstract}
We study analytically and numerically effects of spatial eccentricity of the
 projectile shape on the second flow harmonic in inclusive gluon production
 in p-A collisions in the CGC framework. Keeping the collision area fixed,
 we find that the two quantities are anti-correlated.
\end{abstract}

\maketitle
\section{Introduction}
Currently there is great interest in understanding systematics of correlations in particle production at high energy. Long range in rapidity and angular correlations have long been observed in heavy ion collisions. The current understanding of the origin of these correlations is collective flow in the final state, which in today's most popular incarnation is described in terms of  transport~\cite{transport} and  hydrodynamics~\cite{hydro}. However the observation of very similar correlations in  the p-p and p-Pb collisions at LHC  \cite{Khachatryan:2010gv,Abelev:2012ola,exp-r3,exp-r4,exp-r5}  poses a challenge to hydrodynamic interpretation, at least in small systems. The problem is not only that the final state does not necessarily contain a large number of particle, and that the  correlations extend to relatively high transverse momenta, but also that no jet quenching is observed in these reactions.

It is therefore important to understand better systematic properties of other mechanisms that were proposed as possible origin of correlations at high energy. In particular a considerable amount of work has been done in analyzing the possibility that the observed correlations  do not originate from strong final state interactions, but rather emerge from either the properties of the initial state, or from the initial stages of the collision.   This discussion is most commonly framed in the context of the Color Glass Condensate (CGC), although the physics underlying the effects in question is quite universal and is not necessarily tied to saturation at high energy.

Several potential sources for correlations have been identified in recent literature \cite{correlations}.
In particular in the last couple of years it has been realized that a significant part of the effect discussed so far comes from the quantum interference effects between identical bosons (gluons), as was highlighted in Ref.~\cite{qi}. There are two facets to this effect: the initial state interference - Bose enhancement (BE) of soft gluons in the initial projectile wave function and the final state Hanbury Brown-Twiss (HBT) interference effects between gluons emitted during the early stages of collision. The realizations of the two effects in the spectrum of emitted particles are somewhat different. Both lead to a peak in the number of produced pairs of gluons when the transverse momenta of the two gluons are close to collinear. However the BE  leads to a peak which is broad, and has a width of the typical momentum transfer from the target (target saturation momentum), while the width of the HBT peak is determined primarily by the spatial size of the projectile wave function. As such the relative significance of the two effects in producing correlations is different at different rapidities. The HBT is more important in producing correlations between gluons emitted in the direction of the projectile proton, while the BE should be dominant in the direction of the nucleus.  If the two scales (the saturation momentum of the target and the inverse size of the projectile) are similar, both effects contribute to correlated production.
 In Ref.~\cite{Kovchegov:2018jun},  the role of the quantum interference effects were also identified in the context of
 odd azimuthal harmonics for the two-gluon correlation function in CGC.
The quantum interference effects in gluon production have also been discussed from a somewhat different perspective in Ref.~\cite{urs}.

 So far, with the exception of numerical work of Ref.~\cite{Schenke:2015aqa}, the discussion in the literature has been mostly  semi qualitative. In particular although it is clear that quantum interference effects certainly lead to a nonvanishing $v_2\{2\}$, the dependence of their contribution to $v_2\{2\}$ on the geometry of the collision has not been properly addressed. We note that no correlations between the initial $v_2\{2\}$ and the geometry of the collision have been observed in classical field simulations \cite{Schenke:2015aqa},\cite{Greif:2017bnr} . This question is obviously very important for possible phenomenological applications. The purpose of the present paper is to study this dependence.

 Our interest here is mainly in the collisions between one small and one large object. Consequently the dependence on the geometry in practical terms means the dependence on the spatial shape of the projectile in the transverse plane. When applied to p-A collisions, this translates into dependence of  event by event $v_2\{2\}$ on the configuration of the proton wave function that triggers a given event. We do not endeavor to study geometry fluctuations due to Glauber like fluctuations of the density in the nucleus.

  Our strategy is the following. We will be working entirely within the framework of the dense-dilute CGC approach. We consider inclusive two gluon production from a projectile which  has a nonvanishing spatial eccentricity. This eccentricity is encoded in the distribution of valence sources which produce the soft gluons in the projectile wave function. The averaging over the valence sources is performed using the McLerran-Venugopalan (MV) model modified to incorporate the said spatial eccentricity. We  calculate the ratio of $v_2\{2\}$ corresponding to different eccentricities.  We vary the eccentricity parameter keeping the total area of the projectile, and therefore the single inclusive gluon cross section fixed.

We present both numerical calculations and analytical results. The numerical calculations are performed  without any further approximations. The analytical considerations make  use of the
large $N_c$ approximation and of the parton model like picture which assumes that only the low transverse momentum gluons are present in the projectile wave function.
The analytical and numerical results are consistent with each other and are rather surprising. We find a clear anti correlation between the value of $v_2\{2\}$ and the magnitude of spatial eccentricity.

One might argue that at first sight this trend goes against the observed hierarchy of $v_2\{2\}$ between p-Au and d-Au collisions at RHIC.
However we believe that it would be premature (and thus a mistake) to draw this conclusion, as one also has to take into account the area
dependence, on which current work is missing,  and, most importantly the multiplicity dependence when analyzing the RHIC data.
The dependence on the geometry that we find, although clearly present is rather mild and variation of the
eccentricity from 0 to  0.9 leads only to a minor modification of $v\{2\}$ of order 10\%.
A detailed analysis of p-Au, d-Au and ${ }^3$He-Au collisions is ongoing and will be reported  elsewhere \cite{MSTV}.

  In Section II we explain the general setup of our calculation and the way we introduce the spatial eccentricity.
  In Section III we perform the calculation analytically. Here we use two main approximations. First, we use the target average factorization for products of Wilson lines advocated recently in   Ref.~\cite{qi}. As explained in Ref.~\cite{qi} this approximation is valid in the case of dense target, and it specifically singles out the contributions due to quantum interference. Second, we perform calculations using the projectile source distribution which only allows for the presence of low transverse momentum gluons $q<\Lambda$   in the projectile wave function, where $\Lambda$ is a soft scale. We study production of gluons with  transverse momentum larger than this soft scale $k\gg \Lambda$, but not necessarily larger than the saturation momentum in the target. We consider separately the effect of HBT and BE correlations on $v_2$.

  In Section IV we perform numerical calculation of the same quantity. This time we do not apply large $N_c$ approximation, neither do we introduce the soft scale $\Lambda$, but use the original MV model (modulo eccentricity).

  Section V contains a short discussion of our results.

\section{The setup}
Let us consider the double inclusive gluon production.
We first start with the 
 source-dependent inclusive cross section in the dense-dilute limit 
\begin{equation}
\left.\frac{dN}{d^{2}kdy}\right|_{\rho,U}=\frac{2g^{2}}{(2\pi)^{3}}\int\frac{d^{2}q}{(2\pi)^{2}}\frac{d^{2}q'}{(2\pi)^{2}}\Gamma(\v{k},\v{q},\v{q}')\rho^{a}(-\v{q}')\left[U^{\dagger}(\v{k}-\v{q}')U(\v{k}-\v{q})\right]_{ab}\rho^{b}(\v{q}),
\end{equation}
where the product of two Lipatov vertexes is
\begin{equation}
\Gamma(\v{k},\v{q},\v{q}')=\left(\frac{\v{q}}{q^{2}}-\frac{\v{k}}{k^{2}}\right) \cdot  \left(\frac{\v{q}'}{q'^{2}}-\frac{\v{k}}{k^{2}}\right)\,.
\end{equation}
Here $\rho^a$ is the color charge density in the projectile, and $U$ is the Wilson line in the adjoint representation
in the color field of the target.

The single inclusive and double inclusive production in this approach
are given by
\begin{equation}\label{single}
\left.\frac{dN}{d^{2}kdy}= \left \langle \frac{dN}{d^{2}kdy}\right|_{\rho,U} \right\rangle_{\rho,U}
\end{equation}
and
\begin{equation}\label{double}
\frac{d^{2}N}{d^{2}k_{1}dy_{1}d^{2}k_{2}dy_{2}}= \left \langle \left. \frac{dN}{d^{2}k_{1}dy_{1}}  \right|_{\rho,U}
\left.\frac{dN}{d^{2}k_{2}dy_{2}} \right|_{\rho,U} \right \rangle_{\rho,U}\, ,
\end{equation}
where the brackets denote averaging over the charge density of the projectile and the color fields of the target.

\subsection{Projectile averaging}
We will perform the averaging over the projectile charge density $\rho$ using the MV model, which is equivalent to pairwise Wick contraction of $\rho$ with the basic ``propagator''
\begin{equation}
	\left \langle \rho^a(\v{p}) \rho^b(\v{k}) \right \rangle_\rho
	= \mu^2(\v{p},\v{k}) \delta^{ab}\,.
\end{equation}
In the original MV model the function $\mu^2$ is taken to be proportional to $\delta^2(\v{p}+\v{q})$. This form assumes translational invariance in the transverse plane. Since we wish to explore the dependence on the finite size and shape of the projectile, we generalize it in the following way
\begin{equation}\label{fact}
	\mu^2(\v{p},\v{k}) = \mu^2(\v{p}+\v{k}) F\left(\frac{ (\v{p}-\v{k})^2}{\Lambda^2}\right)\,.
 \end{equation}
 This factorized form albeit not generic, but is intuitive and we believe captures the main features of the projectile charge distribution. The function  $\mu^2(\v{p}+\v{k})$ arises as a Fourier transform of the charge density in the transverse plane
 \begin{equation}
	\mu^2(\v{p})  = \int d^2b e^{i\v{p}\v{b}} \mu^2(\v{b}).
\end{equation}
Thus the coordinate space density profile is directly reflected in $	\mu^2(\v{p})$. Naturally, we expect this function to vanish for momenta much greater than the inverse of the linear dimension of the projectile $R$. The spatial eccentricity will also be directly encoded in $\mu^2$.  The  numerical calculations are performed for a Gaussian profile
\begin{equation}\label{gaus}
	\mu^2(\v{b})  = C e^{-\frac{b_1^2}{a^2 R^2}}  e^{-\frac{a^2 b_2^2}{R^2}}\,.
\end{equation}
The normalization constant $C$ is fixed by
\begin{equation}
	\int d^2b \mu^2(\v{b}) = S_\perp \mu_0^2\, .
\end{equation}
Note that this way of introducing eccentricity {\it preserves} the area of the projectile, and therefore the single inclusive gluon production cross section.

The function $F$  has roughly the meaning of the transverse momentum dependent distribution of the valence partons. In the original MV model one has $F(\v{p})={\rm const}$, corresponding to the point-like structure of the valence color charges. We will use this form in the numerical calculations. For analytic estimates we find it more convenient to assume a partonic picture, according to which the wave function of the projectile contains only low transverse momentum partons. Thus, although we will not use an explicit form of the function $F$, we will assume that it vanishes for momenta larger than some scale $\Lambda$, i.e. $F(x>1)\rightarrow 0$. The scale $\Lambda$ although presumably not hard, has nothing to do with the radius of the projectile. It is likely to be in the range of a single GeV, and thus we will assume that $\Lambda^2\gg 1/R^2$.

 \subsection{Target averaging}
 We also need to average over the field configurations of the target, or equivalently over the Wilson lines $U$. We do this averaging differently in the numerical and analytic calculations.

 The numerical calculations are performed for the physical gauge group $SU(3)$. The averaging over the Wilson lines is performed using the MV model for the target color charge density and calculating the Wilson lines in the resulting ensemble of the target color fields. The details of the procedure are given in Section IV.

 For analytic estimates we rely on the procedure first explained in Ref.~\cite{amir}, and utilized for calculating gluon production in Ref.~\cite{qi3}.
 The average of a product of several  Wilson lines is factorized into contractions formally the same as Wick contractions. The basic ``propagator'' is taken as
\begin{equation}\label{prop}
	\left\langle U_{ab}(\v{p})U_{cd}(\v{q}) \right\rangle_U = \frac{(2\pi)^2}{N_{c}^{2}-1}\delta_{ac}\delta_{bd}\delta^2(\v{p}+\v{q}) D(\v{p});
\end{equation}
where
\begin{equation}
	D(\v{p})=\int d^2xe^{i\v{p}\cdot(\v{x}-\v{y})}\frac{1}{N_c^2-1}\left \langle {\rm tr} \left[ U^\dagger(x)U(y) \right] \right\rangle_U \,.
\end{equation}
This approximation is appropriate for a very dense target and is selecting the terms in the cross section which are not suppressed by powers of area of the projectile~\cite{amir,qi3}. The color structure of the propagator Eq.~(\ref{prop}) reflects the fact that on a dense target the $S$-matrix of a non color singlet state vanishes. Thus the left and right indexes of the Wilson lines have to be separately contracted into a color singlet.

\section{Analytic considerations}
We now return to Eq.~(\ref{double}) and perform the projectile averages. This yields three terms
\begin{eqnarray}\label{rhoav}
&&\frac{d^{2}N}{d^{2}k_{1}dy_{1}d^{2}k_{2}dy_{2}}=\left[\frac{2g^{2}}{(2\pi)^{3}}\right]^2\int
\frac{d^{2}q}{(2\pi)^{2}}\frac{d^{2}q'}{(2\pi)^{2}}\frac{d^{2}p}{(2\pi)^{2}}\frac{d^{2}p'}{(2\pi)^{2}}\Gamma(\v{k}_{1},\v{q},\v{q}')\Gamma(\v{k}_{2},\v{p},\v{p}')\nonumber\\
&&\times\Bigg[\mu^{2}(\v{q}-\v{q}')\mu^{2}(\v{p}-\v{p}')\langle {\rm tr}\left[U^{\dagger}(\v{k}_{1}-\v{q}')U(\v{k}_{1}-\v{q})\right]{\rm tr}\left[U^{\dagger}(\v{k}_{2}-\v{p}')U(\v{k}_{2}-\v{p})\right]\rangle_U\nonumber\\
&&+\mu^2(\v{p} - \v{q}')
\mu^2(\v{q} - \v{p}')
\langle {\rm tr}
[U^{\dagger}(\v{k}_1-\v{q}')U(\v{k}_1-\v{q})
U^{\dagger}(\v{k}_2-\v{p}')U(\v{k}_2-\v{p})]\rangle_U
\\ \notag &&+
\mu^2(-\v{p}' - \v{q}')
\mu^2(\v{q} + \v{p})
\langle {\rm tr}
[U^{\dagger}(\v{k}_1-\v{q}')U(\v{k}_1-\v{q})
U^{\dagger}(\v{p} -  \v{k}_2)U(\v{p}' -  \v{k}_2)]\rangle_U\Bigg] \,.
\end{eqnarray}

\subsection{The double dipole contribution}
First let us consider the first term in Eq.~(\ref{rhoav}). With our target averaging procedure, this yields two type of contributions. The first one where the Wilson line contractions are performed inside each dipole. This yields the square of the single gluon production cross section. It does not  contain correlations, and is not interesting for our purposes.

The other two contributions involve breaking of two color traces, and as shown in Ref.~\cite{qi3} is suppressed by a factor $1/(N_c^2-1)^2$. These terms are subleading at large $N_c$ to other terms that arise from the remaining two lines in Eq.~(\ref{rhoav}), and we will not consider them further.


 Thus we are left to consider the remaining two contractions, which both are suppressed by a single factor $1/(N_c^2-1)$ relative to the uncorrelated double dipole term:
 \begin{align}
&&\left\langle
\overbrace{\rho^{a}(-\v{q}')[U^{\dagger}(\v{k}_1-\v{q}')U(\v{k}_1-\v{q})]_{ab}\rho^{b}(\v{q})}^{dN/d^2k_1dy_1}
\overbrace{\rho^{c}(-\v{p}')[U^{\dagger}(\v{k}_2-\v{p}')U(\v{k}_2-\v{p})]_{cd}\rho^{d}(\v{p})}^{dN/d^2k_2dy_2}
\right\rangle_\rho
\\ \notag && -
\left\langle
\overbrace{\rho^{a}(-\v{q}')[U^{\dagger}(\v{k}_1-\v{q}')U(\v{k}_1-\v{q})]_{ab}\rho^{b}(\v{q})}^{dN/d^2k_1dy_1}
\right\rangle_\rho
\left\langle
\overbrace{\rho^{c}(-\v{p}')[U^{\dagger}(\v{k}_2-\v{p}')U(\v{k}_2-\v{p})]_{cd}\rho^{d}(\v{p})}^{dN/d^2k_2dy_2}
\right\rangle_\rho
\\ \notag  && =
\mu^2(\v{p} - \v{q}')
\mu^2(\v{q} - \v{p}')
{\rm tr}
[U^{\dagger}(\v{k}_1-\v{q}')U(\v{k}_1-\v{q})
U^{\dagger}(\v{k}_2-\v{p}')U(\v{k}_2-\v{p})]
\\ \notag &&+
\mu^2(-\v{p}' - \v{q}')
\mu^2(\v{q} + \v{p})
{\rm tr}
[U^{\dagger}(\v{k}_1-\v{q}')U(\v{k}_1-\v{q})
U^{\dagger}(\v{p} -  \v{k}_2)U(\v{p}' -  \v{k}_2)]\,.
 \end{align}
Each term has two contractions  with respect to $U$ of order $N_c^2$. They comprise the HBT and BE contributions \cite{qi3}.

\subsection{The HBT contribution}
The following contraction leads to the HBT contribution (cyclic property of trace was used)
\begin{align}
&	\mu^2(\v{p} - \v{q}')
\mu^2(\v{q} - \v{p}')
{\rm tr}
[ \left \langle {U(\v{k}_2-\v{p}) U^{\dagger}(\v{k}_1-\v{q}')} \right\rangle_U
\left \langle {U(\v{k}_1-\v{q}) U^{\dagger}(\v{k}_2-\v{p}')} \right\rangle_U ]
\\ \notag &+\mu^2(-\v{p}' - \v{q}')
\mu^2(\v{q} + \v{p})
{\rm tr}
[ \left \langle{U(\v{p}' -  \v{k}_2) U^{\dagger}(\v{k}_1-\v{q}') } \right \rangle_U
\left \langle {U(\v{k}_1-\v{q}) U^{\dagger}(\v{p} -  \v{k}_2)}\right \rangle_U  ]
\\\notag & =  (N_c^2-1)
\mu^2(\v{k}_2 - \v{k}_1)
\mu^2(\v{k}_1 - \v{k}_2)
D(-\v{k}_1+\v{q}')
D(\v{k}_1-\v{q})
\delta(\v{k}_2-\v{k}_1+\v{q}'-\v{p})
\delta(\v{q}'-\v{q}+\v{p}'-\v{p})
\\\notag &+
(N_c^2-1)
\mu^2(\v{k}_1 + \v{k}_1)
\mu^2(-\v{k}_1 - \v{k}_2)
D(-\v{k}_1-\v{q}')
D(\v{k}_1-\v{q})
\delta(\v{k}_1+\v{k}_2-\v{q}-\v{p})
\delta(\v{q}'-\v{q}+\v{p}'-\v{p})\,.
\end{align}
Substituting into double inclusive production we get
\begin{align}
&
\left[ \frac{d^{2}N}
{d^{2}k_{1}dy_{1}d^{2}k_{2}dy_{2}}
\right]_{\rm HBT}
=
(N_c^2-1)
|\mu^2(\v{k}_1 - \v{k}_2)|^2
\left(\frac{2g^{2}}{(2\pi)^{3}}\right)^2
\\ \notag & \times
\int\frac{d^{2}q}{(2\pi)^{2}}\frac{d^{2}p}{(2\pi)^{2}}
\Gamma(\v{k}_1,\v{q},\v{k}_1-\v{k}_2+\v{p})
\Gamma(\v{k}_2,\v{p},\v{k}_2-\v{k}_1+\v{q})D(\v{k}_1-\v{q})
D(\v{p}-\v{k}_2)
+ \left[\v{k}_2 \to - \v{k}_2 \right]\, .
\end{align}
The leading angular dependence in this term is due to the factor $|\mu^2(\v{k}_1 - \v{k}_2)|^2$ which makes the final momentum distribution directly sensitive to the shape of the projectile.

Let us analyze this dependence. In particular we are interested in the second harmonic
\begin{equation}
	\int
	\frac{d\phi_1}{2\pi}
	\frac{d\phi_2}{2\pi}
	e^{2i(\phi_1-\phi_2)}
	\left|
	\int d^2 b e^{i \v{b} \cdot (\v{k}_1-\v{k}_2)}  \mu^2(\v{b})
	\right|^2 \,.
\end{equation}
Analytically we will study the case of  transverse momenta of equal magnitude  $|\v{k}_1| = |\v{k}_2| = k $.

Using the Gaussian density profile Eq.~(\ref{gaus}) and  transforming to the momentum space we find
\begin{equation}
	|\mu^2(\v{k}_1-\v{k}_2)|^2  = (S_\perp \mu_0^2)^2 e^{- (R k)^2
	\left( a^2[\cos \phi_1 - \cos \phi_2 ]^2
+ a^{-2}[\sin \phi_1  - \sin \phi_2  ]^2
   \right)
   }\,.
\end{equation}

We are  interested in the regime of large transverse momenta of observed particles $Rk\gg 1$. We can then expand in $\Delta \phi$, $\phi_1 = \Delta \phi + \phi_2$
and integrate with respect to $\Delta \phi$. In the leading order we get
\begin{align}
&& \int
	\frac{d\phi_1}{2\pi}
	\frac{d\phi_2}{2\pi}
	e^{2i(\phi_1-\phi_2)}
	\left|
	\int d^2 b e^{i \v{b} \cdot (\v{k}_1-\v{k}_2)}  \mu^2(\v{b})
	\right|^2
	= \frac{ (S_\perp \mu_0^2)^2  }{2\pi} \int  \frac{d\phi_2}{2\pi}
	\frac{\sqrt{2 \pi } a}{k R \sqrt{a^4 \sin
   ^2(\phi_2)+\cos ^2(\phi_2)}} +  {\cal O}\left((kR)^0\right)\,.
\end{align}
For small $a$ this integral can be readily calculated with the result
\begin{equation}\label{smalla}
	\lim_{a\to0} \int
	\frac{d\phi_1}{2\pi}
	\frac{d\phi_2}{2\pi}
	e^{2i(\phi_1-\phi_2)}
	\left|
	\int d^2 b e^{i \v{b} \cdot (\v{k}_1-\v{k}_2)}  \mu^2(\v{b})
	\right|^2
	=
	 (S_\perp \mu_0^2)^2
	\frac{8 \sqrt{2 \pi } a \ln
	\left(\frac{2}{a}\right)}{ (2\pi)^2 k R} + {\cal O}\left((kR)^0\right)\,.
\end{equation}
Note that since we introduced the eccentricity parameter $a$ in a way that preserves the  single gluon inclusive production, the integral above is directly proportional to $v_2^2\{2\}$ up to small corrections.

\begin{figure}
\includegraphics[width=0.49\linewidth]{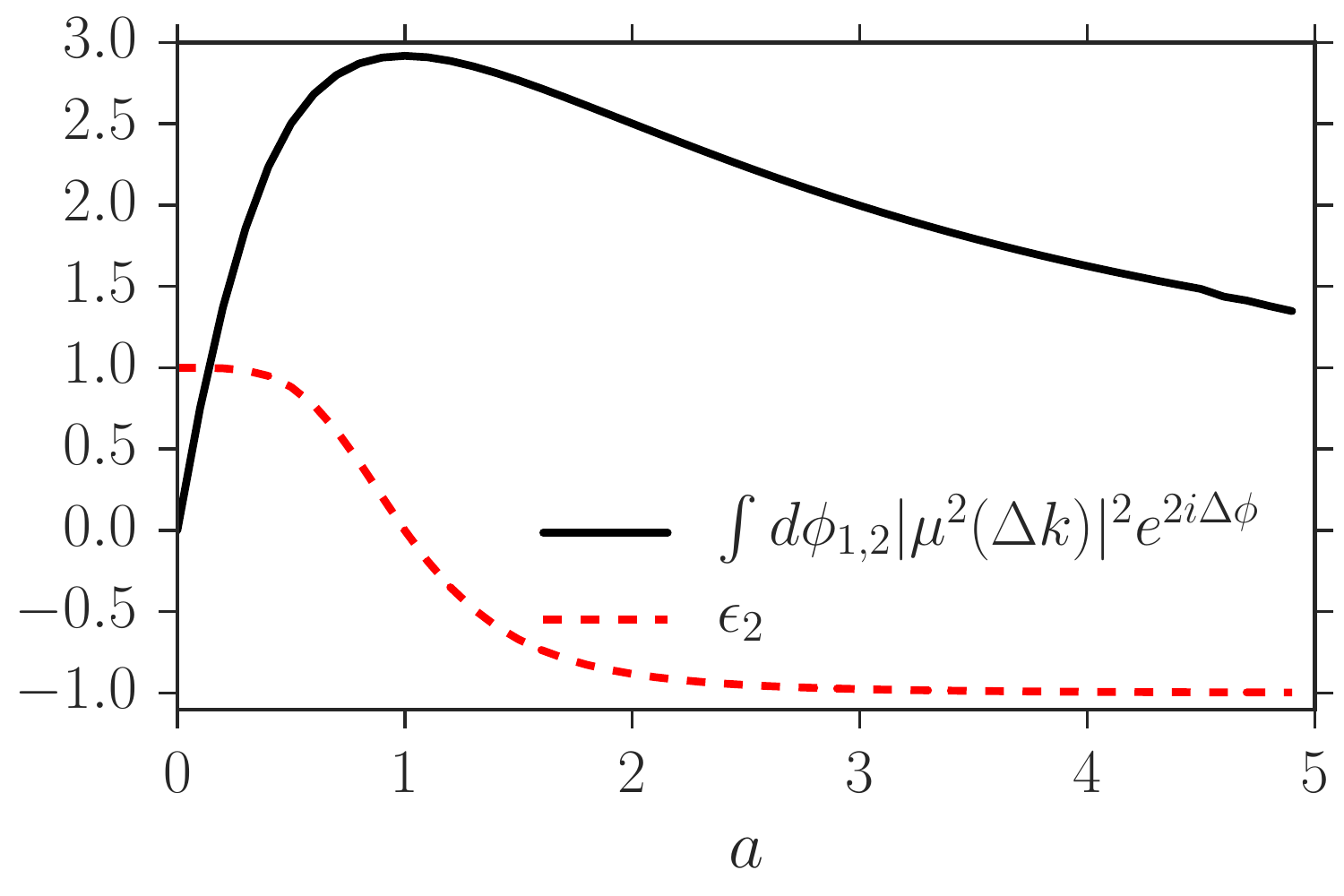}
\includegraphics[width=0.49\linewidth]{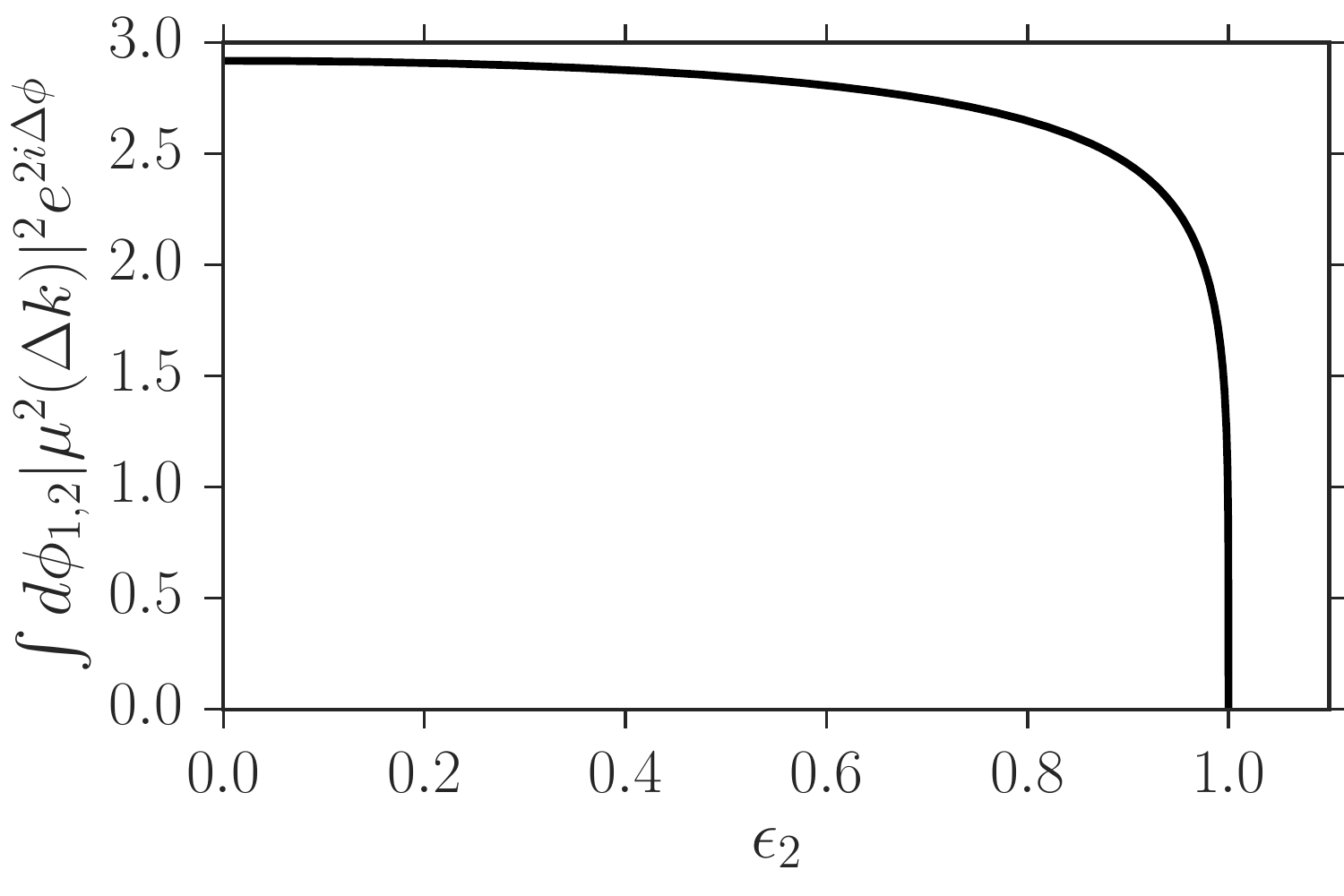}
\caption{$v_2^2\{2\}$ versus spatial eccentricity for the HBT contribution. Left panel: eccentricity and $v_2^2\{2\} $ plotted versus parameter $a$. Right panel $v_2^2 \{2\} $ plotted versus the spatial eccentricity. }
\label{fig:1}
\end{figure}

Thus we find that at least for small $a$ the spatial eccentricity $\epsilon_2$ and $v_2\{2\}$ due to the HBT contribution are anti correlated with each other.
In Fig.~\ref{fig:1} we plot  $v_2^2 \{2\}$ expressed directly in terms of eccentricity for a variety of values of the parameter $a$. We observe the anti correlation for all values of eccentricity. Note, however, that a significant variation of eccentricity
from 0 ($a=1$) to about 0.9 ($a=2$ or $a=1/2$) produces  only  about 10\% change of $\left( \int d\phi_{1,2} |\mu^2(\Delta k)|^{2} e^{2i \delta \phi} \right)^{1/2}$. This mild dependence on the spatial  eccentricity is manifestly demonstrated on the right panel of Fig.~\ref{fig:1}.

In fact although our calculation used explicitly the Gaussian density distribution, the argument is more general. Consider again the relevant quantity
\begin{equation}\label{fint}
\int
	\frac{d\phi_1}{2\pi}
	\frac{d\phi_2}{2\pi}
	e^{2i(\phi_1-\phi_2)}
	\left|
	\int d^2 b e^{i \v{b} \cdot (\v{k}_1-\v{k}_2)}  \mu^2(\v{b})
	\right|^2\,.
	\end{equation}

The Fourier transform limits the difference between the two momenta $|\v{k}_1-\v{k}_2|$  to be of the order of $1/R$. As long as the magnitude of each  momentum is much larger than $1/R$, in the integration region that contributes most significantly to the integral we can write
\begin{equation}
(\v{k}_1-\v{k}_2)_i=\epsilon_{ij}k_j\Delta\phi
\end{equation}
where $\v{k}$ is either of the momenta $\v{k}_i$. This means that the integration over $\Delta\phi$ is effectively limited (by the same phase factor in the Fourier transform) over the range
\begin{equation}
\langle \Delta\phi\rangle\propto\sqrt{ \frac{1}{\langle(\epsilon_{ij}k_ib_j)^2\rangle}}\propto\frac{1}{kR/a}
\end{equation}
where the last relation holds at small $a$. The averages here denote averages over $b$ with the weight given by the density profile $\mu^2(\v{b})$.
Since $\Delta\phi$ is small, we can set the factor $\exp\{i2(\phi_1-\phi_2)\}$ in Eq.~(\ref{fint}) to unity.
It then follows that the integral over $\Delta\phi$ produces a factor of $\frac{a}{kR}$ at small $a$. This is precisely the origin of the explicit factor $a$ in Eq.~(\ref{smalla}).

 This argument is a little naive, since it assumes that the remaining integral is finite in the limit $a\rightarrow 0$. In fact the integral in Eq.~(\ref{smalla}) is logarithmically divergent, which brings about an additional factor of $\ln{2/a}$. This behavior is also quite generic, since
 the expression $\frac{1}{\sqrt{(\epsilon_{ij}k_ib_j)^2}}$ diverges linearly when $\v{b}$ is parallel to $\v{k}$. This divergence  should produce an additional logarithmic dependence on $a$ upon integration over the angle $\phi_2$, since $a$ acts as a regulator for the logarithmic integral. Thus we expect that the result Eq.~(\ref{smalla}) (up to non universal constants) is generically valid for small $a$ and $|\v{k}|\gg 1/R$.

\subsection{The BE contribution}
The remaining contraction over the matrices $U$ in Eq.~(\ref{rhoav}) yields the BE contribution \cite{qi3}:
\begin{align}
&
\mu^2(\v{p} - \v{q}')
\mu^2(\v{q} - \v{p}')
{\rm tr}
[
	\left \langle {U^{\dagger}(\v{k}_1-\v{q}')U(\v{k}_1-\v{q})} \right \rangle_U
\left \langle {U^{\dagger}(\v{k}_2-\v{p}')U(\v{k}_2-\v{p})} \right \rangle_U
 ]
\\ \notag &+
\mu^2(-\v{p}' - \v{q}')
\mu^2(\v{q} + \v{p})
{\rm tr}
[
	\left\langle {U^{\dagger}(\v{k}_1-\v{q}')U(\v{k}_1-\v{q})} \right \rangle_U
\left \langle {U^{\dagger}(\v{p} -  \v{k}_2)U(\v{p}' -  \v{k}_2)} \right \rangle_U  ]
\\ \notag & =
(N_c^2-1)
\mu^2(\v{p} - \v{q}')
\mu^2(\v{q} - \v{p}')
D(-\v{k}_1+\v{q}')
D(-\v{k}_2+\v{p}')
\delta(\v{q}' - \v{q})
\delta(\v{p}' - \v{p})\,.
 \end{align}
Again substituting into double inclusive production  we find
\begin{align}\label{be}
&
\left[ \frac{d^{2}N}
{d^{2}k_{1}dy_{1}d^{2}k_{2}dy_{2}}
\right]_{\rm BE}
=
(N_c^2-1)
\left(\frac{2g^{2}}{(2\pi)^{3}}\right)^2
\\ \notag & \times
\int\frac{d^{2}q}{(2\pi)^{2}}\frac{d^{2}p}{(2\pi)^{2}}
|\mu^2(\v{p} - \v{q})|^2
\Gamma(\v{k}_1,\v{q},\v{q})
\Gamma(\v{k}_2,\v{p},\v{p})
D(\v{q}-\v{k}_1)
D(\v{p}-\v{k}_2)
+ \left[\v{k}_2 \to - \v{k}_2 \right] \,.
\end{align}
Note that  the leading contribution to $\left[ \frac{d^{2}N}
{d^{2}k_{1}dy_{1}d^{2}k_{2}dy_{2}}
\right]_{\rm BE}$ comes from the IR sector of the integrals with respect to
$q$ and $p$, this is due to the presence of the term $1/q^2$  in
\begin{equation}
	\Gamma(\v{k}, \v{q}, \v{q}) =
	\frac{1}{k^2} -
	2 \frac{\v{k}\cdot \v{q}} {k^2 q^2} +
	\frac{1}{q^2} \,.
\end{equation}
This leading contribution however is independent of the angle between $\v{k}_1$ and $\v{k}_2$.

 To analyze the correction to this  contribution we will assume that the momenta $q$ and $p$ are limited in the incoming wave function, as discussed in the previous section Eq.~(\ref{fact}).
In the following we will not indicate the factor $F$ explicitly, but instead will limit the integration over the relevant momentum variable by $\Lambda$.

Then for large $k_i$ we can expand Eq.~(\ref{be}) in  $p$ and $q$. In the following we assume that there is no  preferred direction in the target, so that  $D(\v{k})  = D(|\v{k}|) $. The first term that does not vanish under the angular averaging is
\begin{equation}
|\mu^2(p-q)|^2\frac{(k_1\cdot q)^2(k_2\cdot p)^2}{q^2p^2}\left[\frac{16}{k_1^2k_2^2}D'(k_1^2)D'(k_2^2)+4D''(k_1^2)D''(k_2^2)+\frac{8}{k_1^2}D'(k_1^2)D''(k_2^2)+\frac{8}{k_2^2}D'(k_2^2)D''(k_1^2)\right]\,.
\end{equation}
The factor in the parenthesis does not depend on the angles, and we do not write it out explicitly in the following.

Averaging over the angle we have
\begin{equation}
\Pi_{ij}\equiv\int d\phi e^{2i\phi}\frac{k_ik_j}{k^2}; \ \ \ \Pi_{11}=1/4; \ \ \Pi_{22}=-1/4;\ \ \ \Pi_{12}=i/4\,.
\end{equation}
Thus we are left with calculating the integral
\begin{equation}
\int \frac{d^2p}{(2\pi)^2}\frac{d^2q}{(2\pi)^2} |\mu^2(p-q)|^2[q_x^2-q_y^2+2iq_xq_y][p_x^2-p_y^2-2ip_xp_y]\frac{1}{q^2p^2}=\int d\phi_p d\phi_qe^{2i(\phi_q-\phi_p)} |\mu^2(p-q)|^2 \,.
\end{equation}
Since we are assuming a factorizable ansatz Eq.~(\ref{fact}), it is convenient to define the sum and the difference of the two momenta via
\begin{equation}
\v{q}=\v{Q}+\v{P},\ \  \ \v{p}=\v{Q}-\v{P} \,.
\end{equation}
After some algebra we obtain
\begin{equation}\label{pq}
[q_x^2-q_y^2+2iq_xq_y][p_x^2-p_y^2-2ip_xp_y]\frac{1}{q^2p^2}=1-8\frac{Q_x^2P_y^2+Q_y^2P_x^2}{(Q^2+P^2)^2-4(P\cdot Q)^2}\,.
\end{equation}
As per our assumption the integration over $Q$ extends to significantly larger values than $P$. In both terms in Eq.~(\ref{pq}) the integral over $Q$ is dominated by values much greater than $1/R$. This is true for the first term, where the integral is quadratic at large values of $Q$, but also for the second term where it is logarithmic. Thus with the logarithmic accuracy  in $\ln \Lambda^2R^2$ we can neglect the subleading terms in the denominator in Eq.~(\ref{pq}) and consider the integral
\begin{equation}\label{an}
\int_{Q^2<\Lambda^2} \frac{d^2Q}{(2\pi)^2}\frac{d^2P}{(2\pi)^2} |\mu^2(P)|^2\left[1-8\frac{Q_x^2P_y^2+Q_y^2P_x^2}{(Q^2)^2}\right]=\int_{Q^2<\Lambda^2} \frac{d^2Q}{(2\pi)^2}\frac{d^2P}{(2\pi)^2} |\mu^2(P)|^2\left[1-4\frac{P_y^2+P_x^2}{(Q^2)}\right]\,,
\end{equation}
where the last equality follows since the integration measure over $Q$ is rotationally invariant.
It is now straightforward to extract dependence on the parameter $a$, since $\langle P_x^2 \rangle \propto \frac{1}{a^2R^2}$ and  $\langle P_y^2\rangle \propto \frac{a^2}{R^2}$. Eq.~(\ref{an}) therefore reduces to
\begin{equation}
	v^2_2\{2\}=A-B\left[\frac{1}{a^2}+a^2\right]
\end{equation}
where $A$ and $B$ are positive constants. This is clearly maximal for $a^2=1$ and decreases as a function of spatial eccentricity.

We thus conclude that within our model assumptions both the HBT and BE contributions exhibit anti correlation between the spatial eccentricity of the projectile $\epsilon_2$ and the second flow harmonic in the inclusive gluon production $v_2^2\{2\}$.

In the next section we perform numerical calculations which are not limited by some of the assumptions we had  to make in the present section, in order to check these conclusions in a broader setup.

\section{Numerical results}
For numerical calculations it is easier to work in the following semi-factorizable
representation
\begin{align}
\left.\frac{dN}{d^{2}qdy}\right|_{\rho,U} & =\frac{2}{(2\pi)^{3}}\frac{1}{|\v{q}|^{2}}\left(\delta_{ij}\delta_{lm}+\epsilon_{ij}\epsilon_{lm}\right)\Omega_{ij}^{a}(\v{q})\left[\Omega_{lm}^{a}(\v{q})\right]^{*}\nonumber \\
 & =\frac{2}{2(2\pi)^{3}}\frac{1}{|\v{q}|^{2}}\left(\Omega_{\|}^{a}(\v{q})\left[\Omega_{\|}^{a}(\v{q})\right]^{*}+\Omega_{\perp}^{a}(\v{q})\left[\Omega_{\perp}^{a}(\v{q})\right]^{*}\right)\label{Eq:LO}
\end{align}
where
\begin{equation}
\Omega_{ij}^{a}(\v{x})=g\left[\frac{\partial_{i}}{\partial^{2}}\rho^{b}(\v{x})\right]\partial_{j}U^{ab}(\v{x})\label{Eq:Omega}
\end{equation}
with the adjoint Wilson line defined as
\begin{equation}
U^{ab}(\v{x})=2{\rm tr}\left[t^{b}V^{\dagger}(\v{x})t^{a}V(\v{x})\right]\label{Eq:Adj}
\end{equation}
and
\begin{align}
\Omega_{\|}^{a}(\v{k}) & =\delta_{lm}\Omega_{lm}^{a}(\v{k})\equiv\Omega_{11}^{a}(\v{k})+\Omega_{22}^{a}(\v{k})\,,\\
\Omega_{\perp}^{a}(\v{k}) & =\epsilon_{lm}\Omega_{lm}^{a}(\v{k})\equiv\Omega_{12}^{a}(\v{k})-\Omega_{21}^{a}(\v{k})\,.
\end{align}

Thus to compute $2$-particle inclusive production we have to
\begin{itemize}
	\item[--] Generate an ensemble of configurations of the projectile charge density $\rho$ and the analogously of the color charge density in the target $\rho_{t}$ on a two-dimensional lattice. We generate the two ensembles using the MV model with parameters $\mu^2$ and $\mu^2_t$ respectively.
	\item[--] Compute corresponding fields $\alpha=\frac{1}{\partial^{2}}\rho$
by using standard Fourier transform method
\item[--] Evaluate the fundamental Wilson line $V(x)$ and the adjoint Wilson
line $U(x)$ for the target
\item[--] Using finite (central) difference scheme compute $\partial_{i}\alpha$
and $\partial_{i}U(x)$
\item[--] Combine the results into $\Omega_{\|}^{a}(\v{x})$ and $\Omega_{\perp}^{a}$
\item[--] Compute Fourier Transform  for $\Omega_{\|,\perp}^{a}(\v{x})\to\Omega_{\|,\perp}^{a}(\v{k})$
\item[--] Combine to
\begin{equation}
\left.\frac{dN}{d^{2}qdy}\right|_{\rho,U}=\frac{2}{(2\pi)^{3}}\frac{1}{|\v{q}|^{2}}\left(\Omega_{\|}^{a}(\v{q})\left[\Omega_{\|}^{a}(\v{q})\right]^{*}+\Omega_{\perp}^{a}(\v{q})\left[\Omega_{\perp}^{a}(\v{q})\right]^{*}\right)
\end{equation}
\item[--] Compute $2$-gluon inclusive production.
\end{itemize}
We have performed the numerical calculation for two values of $\mu/\mu_{\rm t}$ and different values of observed momenta.
We fixed  the radius of the projectile $R=\mu^{-1}$. More details on the numerical implementation can be found in
Refs.~\cite{Lappi:2007ku,Dumitru:2014vka}.

\begin{figure}
\centerline{\includegraphics[width=0.49\linewidth]{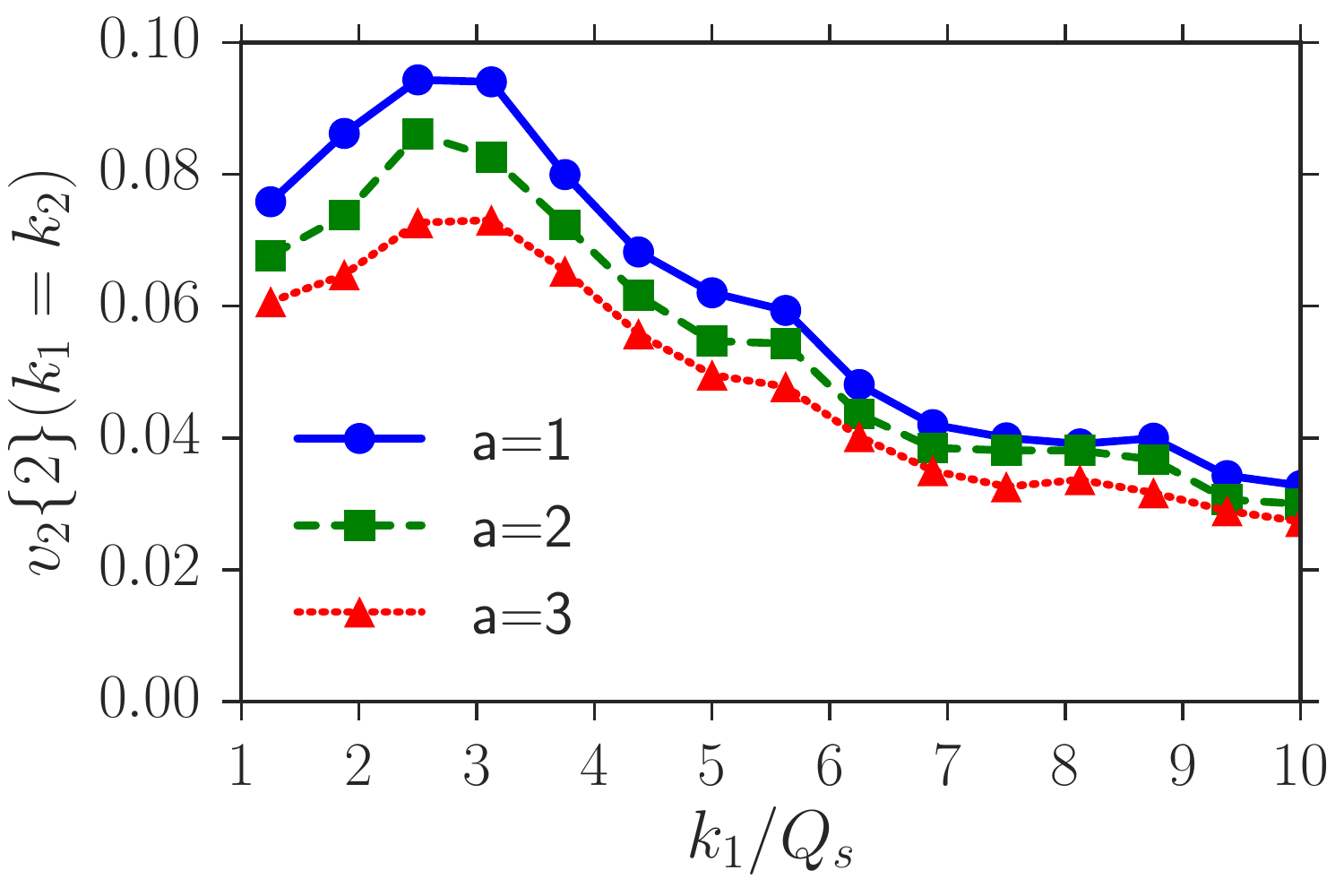}
\includegraphics[width=0.49\linewidth]{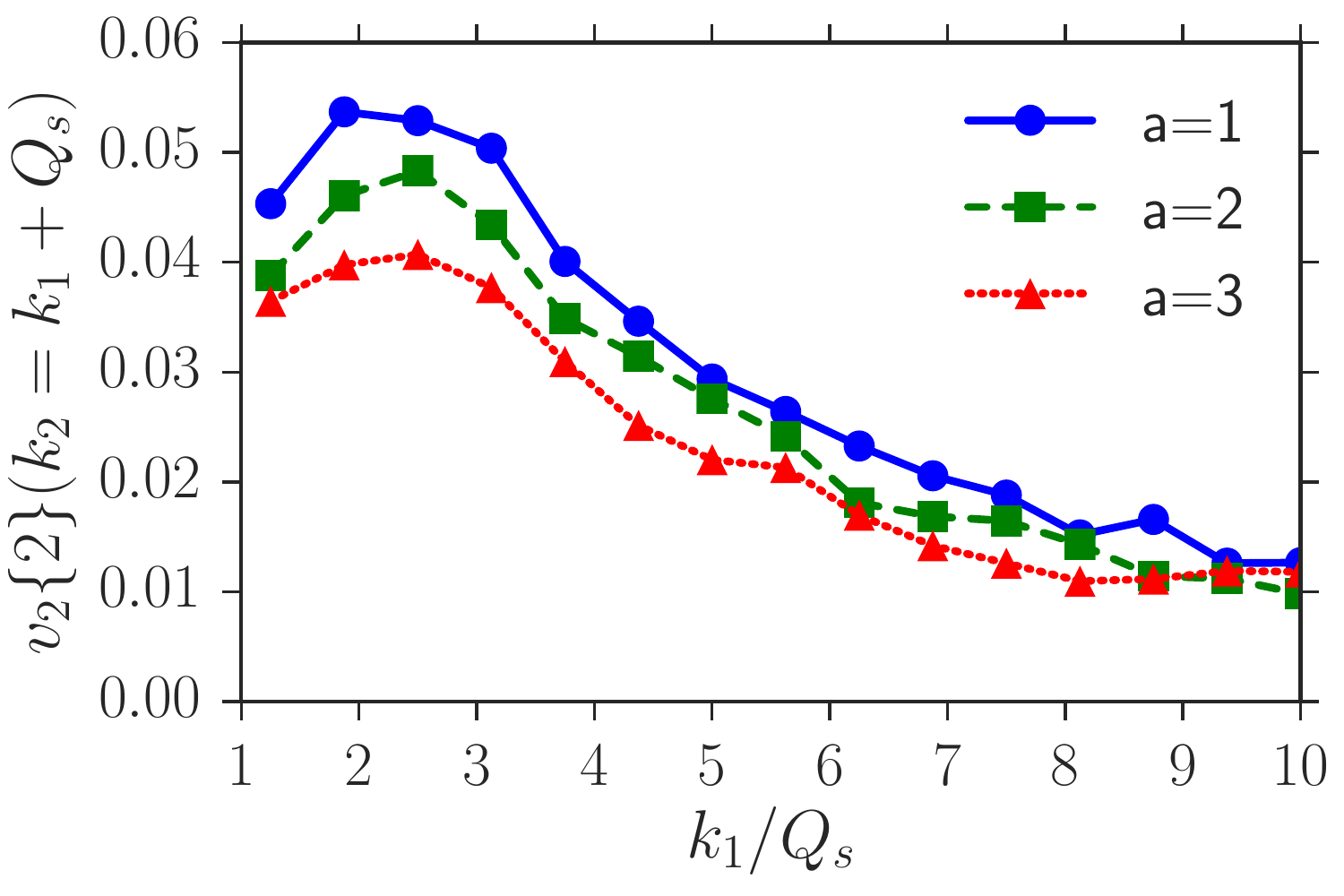}}
\caption{$v_2\{2\}$ as a function of momentum and the anisotropy of the projectile $a$. The total area of the projectile
is kept independent of $a$. The left panel: both gluons are at the same absolute value of momenta $k_1=k_2$.
The right panel: the difference between the momenta of gluons is given by the saturation momentum of
the target; this excludes HBT contribution.  $\mu = \mu_{\rm t}/4$.
}
\label{fig:mu025}
\end{figure}

\begin{figure}
	\centerline{\includegraphics[width=0.49\linewidth]{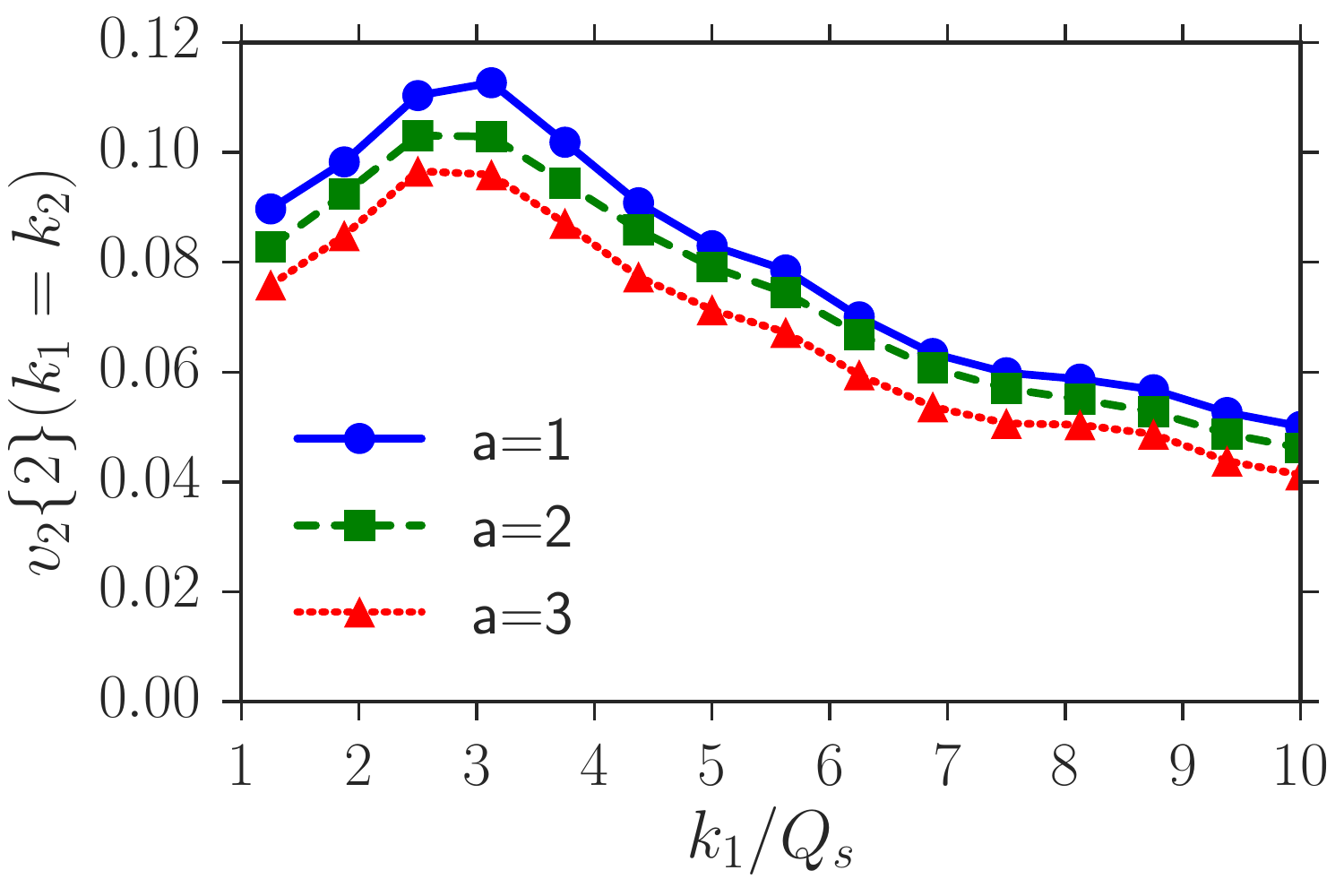}
	\includegraphics[width=0.49\linewidth]{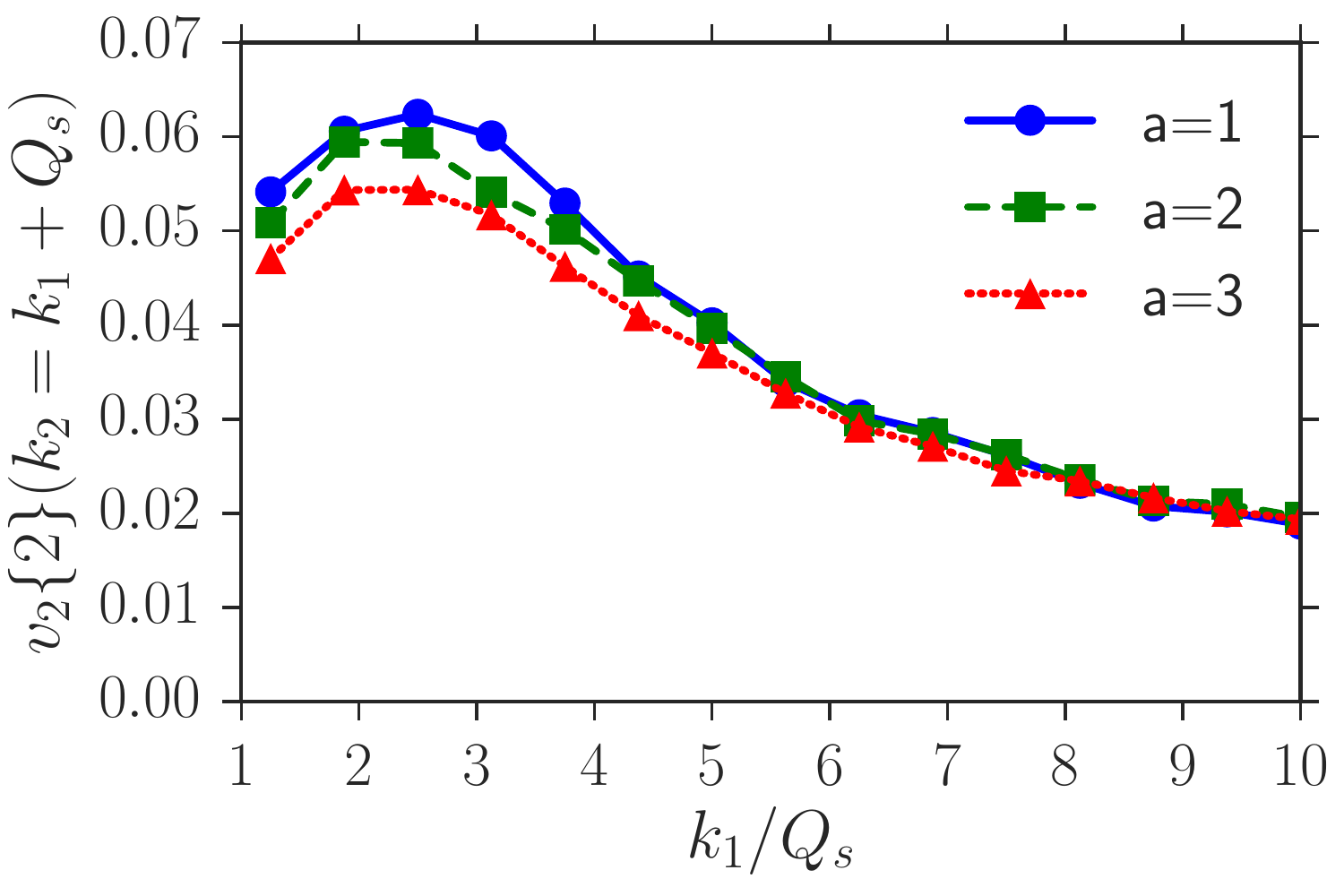}}
\caption{
	The same as in Fig.~\ref{fig:mu025}, but for   $\mu = \mu_{\rm t}/2$.
}
\label{fig:mu05}
\end{figure}

Fig.~\ref{fig:mu025} displays the results for $\mu/\mu_{\rm t}=1/4$. The left panel shows $v_2$ for different values of the eccentricity parameter $a$ as a function of momentum. On the left panel the magnitude of the momenta of the two gluons are the same. In this regime we expect the correlation to be dominated by the HBT effect. On the right panel the difference of the magnitudes is the saturation momentum of the target $Q_s$.  We expect this regime to be outside the narrow HBT peak, and the correlation to be dominated by the Bose enhancement. In all cases we observe consistent anti correlation between $v_2\{2\}$ and the spatial eccentricity. Fig.~\ref{fig:mu05} displays the same for $\mu/\mu_{\rm t} =1/2$. The results qualitatively are very similar.

\section{Discussion}
In this paper we have studied the dependence of the second flow harmonic $v_2\{2\}$ in inclusive gluon production on the spatial eccentricity of the projectile within the CGC approach. We have performed analytic estimates and numerical calculations. Our analytic considerations assume parton model like distribution of valence charges in the projectile, and explicitly concentrate on the HBT and Bose enhancement contributions to correlated gluon production. Numerical calculations use the original MV model, which contains a perturbative hard tail  of valence gluons, and treat all contributions to correlated production on an equal basis.

In all the analytic and numerical calculations we find consistently that the magnitude of the second flow harmonic is anti correlated with the spatial eccentricity of the projectile.

We note that dependence of the second flow harmonic on geometry of a collision in the CGC approach was studied in Ref.~\cite{doug} in a different physical situation. The authors of Ref.~\cite{doug} analyzed the dependence of $v_2\{2\}$ on relative spatial orientation of two ellipsoidal colliding objects (``uranium nuclei''). They  concluded that for a symmetric ``tip on'' collision $v_2\{2\}$   was larger than for an eccentric ``side on'' configuration. This conclusion is qualitatively similar to ours, although 
the two analysis probe different physics. Changing the relative orientation of colliding objects changes the collision area, which in the CGC
approach immediately leads to variation in $v_2\{2\}$ in a way that does not depend on momentum of observed particles.
This 
 is the origin of the effect observed in Ref.~\cite{doug}. On the other hand our analysis is performed varying eccentricity, but keeping the interaction area fixed. To our knowledge such an analysis has not been performed before, and in fact it was believed that in such situation there should be no effect of geometry on $v_2\{2\}$, see e.g. Refs.~\cite{Schenke:2015aqa,Greif:2017bnr}. Note, however, that in order to be able to
 detect the effect of  geometry numerically we had to consider rather large  eccentricities, $a=2$ ($|\epsilon_2| \approx 0.88$)  and $3$
 ($|\epsilon_2| \approx 0.98$). Changing $a$ from $1$ (zero eccentricity)  to
 $2$, leads only to about 10\% variation in  $v_2\{2\}$, as the
 bulk contribution to $v_2\{2\}$ is  geometry independent.

It is clear that this effect must have some phenomenological implications. At first sight this trend goes against the observed hierarchy of $v_2$ between p-Au and d-Au collisions at RHIC. However, based on the above argument,  we believe that it is too early to draw conclusions, as one also has to take into account the area dependence and, most importantly the multiplicity dependence when analyzing the RHIC data. A detailed analysis of p-Au, d-Au and  ${ }^3$He-Au
collisions is ongoing and will be reported
elsewhere \cite{MSTV}.

\begin{acknowledgments}
We thank
Adam Bzdak,
Adrian Dumitru,
Michael Lublinsky,
Larry McLerran,
Prithwish  Tribedy,
Urs Wiedemann
for  useful discussions on problems related to this project.

V.S. is grateful to Yury Kovchegov, Mark Mace and  Raju Venugopalan for illuminating conversations.

This research was supported by the NSF Nuclear Theory grant 1614640  and CERN scientific associateship (A.K.).
V.S. is indebted to Urs Wiedemann and CERN TH group for the partial support and hospitality at CERN, where this work was finalized.
V.S. also gratefully acknowledges partial support by the ExtreMe Matter Institute EMMI
(GSI Helmholtzzentrum f\"ur Schwerionenforschung, Darmstadt, Germany).

\end{acknowledgments}

\end{document}